\documentclass[a4paper]{SIGEport}
\PassOptionsToPackage{table}{xcolor}
\usepackage{graphicx}
\usepackage{geometry}
\usepackage{cite}
\usepackage{amsmath}
\usepackage{booktabs}
\usepackage{multirow}
\usepackage{array}
\usepackage{arena-style}

\newcommand{\ArenaLLMModel}{Claude Opus 4.8}
\newcommand{\ArenaLLMModelId}{\texttt{claude-opus-4-8}}
\newcommand{\ArenaLLMElasticDet}{1.00}
\newcommand{\ArenaLLMElasticGrnd}{1.00}

\newcommand{\ArenaLLMSplunkDet}{1.00}
\newcommand{\ArenaLLMSplunkGrnd}{1.00}

\newcommand{\ArenaLLMOsqueryDet}{0.20}
\newcommand{\ArenaLLMOsqueryGrnd}{0.20}

\newcommand{\ArenaLLMOsqueryBE}{1}

\makeatletter
\def\resumo{\normalfont%
  \if@twocolumn%
    \@IEEEabskeysecsize\bfseries\textit{Abstract}---\,%
  \else%
    \begin{center}\vspace{-1.78ex}\@IEEEabskeysecsize\textbf{Abstract}\end{center}\quotation\@IEEEabskeysecsize%
  \fi\@IEEEgobbleleadPARNLSP}
\def\chave{\normalfont%
  \if@twocolumn%
    \@IEEEabskeysecsize\bfseries\textit{Keywords}---\,\relax%
  \else%
    \begin{center}\vspace{-1.78ex}\@IEEEabskeysecsize\bfseries Keywords\end{center}\quotation\@IEEEabskeysecsize%
  \fi\@IEEEgobbleleadPARNLSP}
\def\fnum@table{\scriptsize{TABLE~\thetable}}

\def\thebibliography#1{\section*{References}%
  \addcontentsline{toc}{section}{References}%
  \footnotesize \vskip 0.3\baselineskip plus 0.1\baselineskip minus 0.1\baselineskip%
  \list{\@biblabel{\@arabic\c@enumiv}}%
  {\settowidth\labelwidth{\@biblabel{#1}}%
  \leftmargin\labelwidth
  \advance\leftmargin\labelsep\relax
  \itemsep 0pt plus .5pt\relax%
  \usecounter{enumiv}%
  \let\p@enumiv\@empty
  \renewcommand\theenumiv{\@arabic\c@enumiv}}%
  \let\@IEEElatexbibitem\bibitem%
  \def\bibitem{\@IEEElatexbibitem}%
  \def\newblock{\hskip .11em plus .33em minus .07em}%
  \if@technote\sloppy\clubpenalty4000\widowpenalty4000\interlinepenalty100%
  \else\sloppy\clubpenalty4000\widowpenalty4000\interlinepenalty500\fi%
  \sfcode`\.=1000\relax}

\makeatother

\usepackage[colorlinks=true,linkcolor=black,citecolor={blue!45!black},urlcolor={blue!45!black}]{hyperref}

\newif\ifanonymous
\anonymoustrue
\ifdefined\anonymous
  \ifnum\anonymous=0
    \anonymousfalse
  \fi
\fi

\begin{document}

\title{ARENA: An Architecture for Measuring the Transferability of Autonomous Cyber Defense}

\author{Sidnei Barbieri$ ^1 $, \'{A}gney Lopes Roth Ferraz$ ^1 $, Wagner Comin Sonaglio$ ^1 $, Gioliano de Oliveira Braga$ ^1 $, Henrique Curi de Miranda$ ^1 $ and Louren\c{c}o Alves Pereira J\'{u}nior$ ^1 $\\
    {\small $ ^1 $Computer Science Division, Aeronautics Institute of Technology (ITA), S\~{a}o Jos\'{e} dos Campos/SP - Brazil}\\
    {\small sidneisb@ita.br, roth@ita.br, sonaglio@ita.br, giolianobraga@ita.br, henriquecuri1@proton.me, ljr@ita.br}}

\maketitle

\begin{resumo}
Agentic AI systems powered by Large Language Models (LLMs) are increasingly used to build autonomous defense agents, yet evaluating them remains an open challenge. An agent that succeeds in one environment often degrades when the same attack runs on a different platform, telemetry stack, or policy; we call this the \textit{agent transferability gap}: success in one environment is not evidence of capability in the next. We present ARENA, a modular architecture that decomposes cyber-defense evaluation into four independently varying layers: attacker emulation, System Under Test (SUT) generation, a typed agent runtime, and a deterministic policy verifier, plus a failure taxonomy that attributes degradation to schema, grounding, budget, or policy. A controlled experiment reveals the gap: two deterministic agents, indistinguishable under detection recall, diverge once the telemetry schema changes, and ARENA attributes the loss to observation semantics rather than to attack recognition. ARENA makes transferability measurable, attributable, and reproducible.
\end{resumo}

\begin{chave}
autonomous agents, transferability, cyber range, adversary emulation
\end{chave}

\section{Introduction}

Security Operations Centers (SOCs) process thousands of alerts a day, yet the median time to contain a breach is still measured in days, not minutes \cite{ibmcost2024}. That pressure is moving Large Language Model (LLM) agents into the loop as autonomous defenders that triage, investigate, and contain with little human intervention \cite{pocketagents}. Each such agent is validated somewhere, on a benchmark, in a lab, or on one cyber range, and is then deployed elsewhere. Deployment rests on an assumption that no one states: that a capability shown in one environment carries over to the next.

The assumption is fragile. The same attack produces different observations as the environment changes: Elastic Common Schema (ECS) JavaScript Object Notation (JSON) becomes Splunk Common Information Model (CIM) records, a Windows event log becomes a Linux audit record, and an enterprise host becomes an Open Radio Access Network (O-RAN) controller. An agent that learned to read one telemetry schema can miss the identical attack expressed in another. When it does, current evaluations remain silent on the questions that determine whether the agent is safe to deploy: whether its capability degraded, by how much, at which step in its reasoning, and why.

So this paper puts one question first: \textit{can an autonomous cyber-defense capability transfer across environments?} The difficulty is structural. Structured Cyber Threat Intelligence (CTI), such as MITRE ATT\&CK encoded in the Structured Threat Information Expression (STIX) format, constrains but does not uniquely determine the target environment \cite{autosut}: one campaign can be faithfully realized on Linux containers, Windows virtual machines, or a hybrid cloud, each emitting different telemetry. The attack intent is fixed; the environment and the observations it exposes are not. We call the resulting capability loss the \textit{agent transferability gap}, and current practice cannot see it. Static prompt benchmarks divorced from infrastructure \cite{secure2024} and single-environment ranges both ask whether an agent worked in one context; neither asks whether its competence survives a change of context.

Before ARENA, transferability is assumed. An evaluator who watches an agent fail can report that it failed, not whether the failure came from a changed telemetry schema, a lost entity grounding, an exhausted query budget, or a policy the agent misread. After ARENA, transferability is measured, attributed, and reproduced: ARENA holds the attack intent fixed, varies the environment, and records enough of each run to replay it, so a degradation becomes a number with a cause rather than an anecdote.

ARENA (Agent Resilience Evaluation Architecture) contributes a decomposition of cyber-range evaluation into four independently variable subsystems: attacker emulation, System Under Test (SUT) generation, agent runtime, and policy verification; a failure taxonomy that attributes degradation to schema, grounding, budget, or policy; and a controlled, reproducible experiment that isolates transferability failures across platform variants. Enterprise IT is the primary validation case. Artificial Intelligence Radio Access Network (AI-RAN) over O-RAN is a deliberate stress case: it changes the environment and observation semantics more aggressively than any enterprise variant, motivating a transferability architecture rather than a single-domain benchmark.

\section{Problem Formulation}
\label{sec:problem}

\begin{figure}[t]
\centering
\begin{tikzpicture}[arena, font=\footnotesize]
  \node[arena node, text width=3.0cm] (eA)
    {\textbf{Environment A}\\\scriptsize correlated telemetry\\(Elastic, Splunk)};
  \node[arena node, right=10mm of eA, text width=3.0cm] (eB)
    {\textbf{Environment B}\\\scriptsize normalized telemetry\\(osquery)};
  \coordinate (etop) at ($(eA.north)!0.5!(eB.north)$);
  \node[arena attack, above=10mm of etop, text width=5.2cm] (top)
    {\textbf{Same attack, same defense agent}};
  \coordinate (fY) at ($(top.south)+(0,-3mm)$);
  \draw[arena flow strong, -] (top.south) -- (fY);
  \draw[arena flow strong, -] (eA.north|-fY) -- (eB.north|-fY);
  \draw[arena flow strong] (eA.north|-fY) -- (eA.north);
  \draw[arena flow strong] (eB.north|-fY) -- (eB.north);
  \node[arena policy, below=7mm of eA, text width=3.0cm] (oA)
    {\textbf{succeeds}};
  \node[arena box, draw=arenaRed, fill=arenaRedBg, below=7mm of eB, text width=3.0cm] (oB)
    {\textbf{fails}};
  \draw[arena flow strong] (eA.south) -- (oA.north);
  \draw[arena flow strong] (eB.south) -- (oB.north);
  \coordinate (obot) at ($(oA.south)!0.5!(oB.south)$);
  \node[arena box, draw=arenaInk, below=10mm of obot, text width=6.4cm] (why)
    {Only the telemetry schema changed.\\\textbf{Why different outcomes?}};
  \coordinate (mY) at ($(oA.south)+(0,-3mm)$);
  \draw[arena flow strong, -] (oA.south) -- (oA.south|-mY);
  \draw[arena flow strong, -] (oB.south) -- (oB.south|-mY);
  \draw[arena flow strong, -] (oA.south|-mY) -- (oB.south|-mY);
  \draw[arena flow strong] ($(oA.south|-mY)!0.5!(oB.south|-mY)$) -- (why.north);
  \node[arena policy, below=7mm of why, text width=6.4cm] (arena)
    {\textbf{ARENA} measures the gap, attributes a cause, and makes it reproducible.};
  \draw[arena flow strong] (why.south) -- (arena.north);
\end{tikzpicture}
\caption{The agent transferability gap.}
\label{fig:problem}
\end{figure}

To measure transferability, we must name what is held fixed and what is allowed to change between two evaluations of the same agent. Informally, every evaluation couples five things: the attack the agent faces, the environment it runs in, what that environment lets it see, the agent itself, and the rules it must obey. Today, these are entangled, so when an agent fails, no one can say which of them caused it. Naming them lets us hold some fixed and vary others, which is exactly what current evaluation cannot do.

We make the five precise as a tuple $\langle A, E, O, \pi, P\rangle$. The \textit{attack intent} $A$ is the behavioral semantics of a campaign expressed in structured CTI (ATT\&CK or FiGHT techniques in STIX), independent of any concrete host. The \textit{environment} $E$ is a concrete System Under Test: operating system, services, topology, and telemetry stack. The \textit{observations} $O$ are the telemetry that the environment exposes to the defender; the same intent on two environments yields different observations, because ECS JSON, Splunk CIM records, and O-RAN E2 Service Model (E2SM) counters encode the same events in different schemas. The \textit{agent} $\pi$ is a defense policy mapping observations $O$ to defensive actions, and the \textit{operational policy} $P$ is the governance constraints an action must respect.

A defender's competence is therefore a property of the whole tuple, not of $\pi$ alone. Holding the intent $A$ and the governance $P$ fixed, we vary the environment from $e_i$ to $e_j$, thereby varying the observations it exposes. \textit{Transferability} is the degree to which the competence of $\pi$ survives that shift, read per exercised dimension: detection, grounding, resource cost, policy compliance, and, for live SUTs, time-to-containment. A result that holds for $e_i$ and collapses to $e_j$ is the agent transferability gap, made quantitative (Fig.~\ref{fig:problem}).

This framing reorients evaluation. A static benchmark fixes $E$ and $O$ and measures $\pi$ on prompts; a single-environment range fixes one $E$ and varies $A$. Neither isolates the variable that matters in deployment, the environment, so neither can separate a capability intrinsic to $\pi$ from an artifact of the one environment it was tuned on. ARENA makes $\langle A, E, O, \pi, P\rangle$ explicit, varies $E$ and $O$ while holding $A$ and $P$ fixed, and attributes each change in transferability to the layer that produced it. The rest of the paper instantiates each element of the tuple as an independently configurable layer and defines the taxonomy that performs the attribution.

\textbf{Threat model and scope.} ARENA targets autonomous defense against targeted, multi-stage campaigns, the behavior of an Advanced Persistent Threat (APT) rather than an opportunistic scan or a single exploit. The attack intent $A$ is therefore drawn from documented campaigns encoded in structured CTI: enterprise campaigns from ATT\&CK-in-STIX and, in the AI-RAN domain, telecom campaigns from MITRE FiGHT. The empirical claim is precise: ARENA quantifies whether a defender's competence survives the environmental changes a campaign traverses and attributes any loss to a layer. The adversary follows the campaign's behavioral semantics while the environment fixes concrete bindings, such as hosts, credentials, and telemetry schema, making one intent replayable across SUTs. The evaluation boundary is the deployed agent and its enforcement mechanism; the physical layer, model-training supply chain, and human analyst remain separate objects of study.

\section{Background}

Cyber ranges have evolved from static training environments for human operators into programmable platforms capable of automated adversary emulation. Tools such as Apache Caldera~\cite{caldera2023}, originally developed at MITRE, orchestrate multi-step attack campaigns by executing ``abilities'' mapped to ATT\&CK techniques. However, Caldera and similar platforms assume that the operator manually configures the target environment and provides the specific parameters (IP addresses, credentials, file paths) required to map abstract techniques to executable commands.

Recent measurement work has quantified this gap. The \textit{procedural semantics gap} in ATT\&CK-in-STIX reveals that structured technique-to-procedure entries lack the executable detail needed for fully automated replay \cite{sticks}. Meanwhile, the \textit{environment semantics gap} demonstrates that the overwhelming majority of software objects in public CTI lack version or Common Platform Enumeration (CPE) pinning, making automated SUT derivation impossible without analyst intervention \cite{autosut}.

The integration of LLMs into security operations has progressed along two tracks. \textit{Copilot} systems, such as Microsoft Security Copilot, augment human analysts by summarizing alerts and suggesting investigation steps. \textit{Agent} systems go further: they execute closed-loop investigation and containment without waiting for human approval at each step \cite{pocketagents}.

Both tracks introduce failure modes absent from traditional rule-based systems. LLMs can hallucinate IP addresses that are not present in the telemetry, propose actions against assets outside their jurisdiction, or violate organizational policies, such as isolating a production server before collecting forensic evidence \cite{socpilot}.

Static benchmarks test prompt-level reasoning but ignore infrastructure coupling. Single-environment simulations capture interactions within a fixed topology. Agent guardrails usually check only the prompt or tool boundary. None varies the environment $E$ and the observations $O$ under a fixed intent $A$, which is the regime in which the transferability question of Section~\ref{sec:problem} lives.

\section{ARENA Architecture}

ARENA decomposes the evaluation problem into four subsystems, each independently configurable. Fig.~\ref{fig:architecture} shows them with explicit interfaces over a shared artifact store: the attack intent enters the attacker substrate, the SUT generator instantiates the environment, the agent runtime emits a typed action trace, and the policy verifier checks it. Every subsystem reads and writes versioned artifacts, so any run can be replayed or audited, and a transferability failure can be traced to the subsystem that produced it.

\begin{figure*}[t]
\centering
\begin{tikzpicture}[arena, font=\scriptsize, node distance=2mm and 12mm]
  \node[arena attack, text width=3.0cm] (a1) {CTI / STIX bundle $\to$ intent model};
  \node[arena attack, text width=3.0cm, below=of a1] (a2) {Parameter binding $\to$ Caldera adapter};
  \node[arena node, text width=3.0cm, right=13mm of a1.east, anchor=west] (b1) {Fixed $\cap$ free region};
  \node[arena node, text width=3.0cm, below=of b1] (b2) {Provisioning orchestrator};
  \node[arena node, text width=3.0cm, below=of b2] (b3) {SUT-A/B/C \\ Win$\cdot$Elastic, Win$\cdot$Splunk, Linux$\cdot$osquery};
  \node[arena node, text width=3.0cm, right=13mm of b1.east, anchor=west] (c1) {Dispatcher $\to$ agent (LLM) + tools};
  \node[arena node, text width=3.0cm, below=of c1] (c2) {Typed schema + grounding boundary};
  \node[arena policy, text width=3.0cm, right=13mm of c1.east, anchor=west] (d1) {task $\mid$ action $\mid$ source $\mid$ data};
  \node[arena policy, text width=3.0cm, below=of d1] (d2) {Deterministic verifier $\to$ verdict};
  \coordinate (gbot) at (b3.south);
  \begin{scope}[on background layer]
    \node[arena group, fit=(a1)(a2)(a1|-gbot)] (S1){};
    \node[arena group, fit=(b1)(b2)(b3)]        (S2){};
    \node[arena group, fit=(c1)(c2)(c1|-gbot)] (S3){};
    \node[arena group, fit=(d1)(d2)(d1|-gbot)] (S4){};
  \end{scope}
  \node[font=\footnotesize\bfseries, anchor=south] at (S1.north) {1.\ Attacker emulation};
  \node[font=\footnotesize\bfseries, anchor=south] at (S2.north) {2.\ SUT generator};
  \node[font=\footnotesize\bfseries, anchor=south] at (S3.north) {3.\ Agent runtime};
  \node[font=\footnotesize\bfseries, anchor=south] at (S4.north) {4.\ Policy verifier};
  \draw[arena flow strong] (S1.east|-a2) -- (S2.west|-a2);
  \draw[arena flow strong] (S2.east|-a2) -- (S3.west|-a2);
  \draw[arena flow strong] (S3.east|-a2) -- (S4.west|-a2);
  \coordinate (bl) at (S1.south west);
  \coordinate (br) at (S4.south east);
  \coordinate (fc) at ($(bl)!0.5!(br)$);
  \node[arena box, draw=arenaInk, dashed, text width=17.2cm, anchor=north]
    at ($(fc) + (0,-9mm)$) (store)
    {\textbf{Evaluation engine \& versioned artifact store}\quad\textcolor{arenaGray}{transferability matrix · failure taxonomy (SF/GF/BE/PV) · immutable, replayable runs}};
  \coordinate (interfaceY) at ($(S2.south)!0.5!(store.north)$);
  \coordinate (iface12) at ($(S1.east)!0.5!(S2.west)$);
  \coordinate (iface23) at ($(S2.east)!0.5!(S3.west)$);
  \coordinate (iface34) at ($(S3.east)!0.5!(S4.west)$);
  \node[arena label, text=arenaInk] at (iface12|-interfaceY) {\bfseries intent $A$};
  \node[arena label, text=arenaInk] at (iface23|-interfaceY) {\bfseries observations $O$};
  \node[arena label, text=arenaInk] at (iface34|-interfaceY) {\bfseries action trace};
  \draw[arena flow, dashed] (S1.south) -- (S1.south|-store.north);
  \draw[arena flow, dashed] (S2.south) -- (S2.south|-store.north);
  \draw[arena flow, dashed] (S3.south) -- (S3.south|-store.north);
  \draw[arena flow, dashed] (S4.south) -- (S4.south|-store.north);
\end{tikzpicture}
\caption{ARENA measurement architecture. The shared store records a transferability matrix and a failure taxonomy: schema (SF), grounding (GF), budget (BE), and policy (PV) failures.}
\label{fig:architecture}
\end{figure*}

\textbf{Attacker substrate.} The first layer converts structured CTI into executable, multi-host attack workflows. It separates behavioral extraction, which parses ATT\&CK technique IDs and STIX procedure descriptions into adversary intent, from parameter binding, which maps that intent to target IPs, credentials, paths, and other SUT-specific values, and from workflow compilation, which emits a plan for Caldera or an equivalent orchestrator~\cite{sticks}. This separation is what makes the transferability testable: lateral movement over a Server Message Block (SMB)-class share reflects the same attack intent whether it runs on Windows SMB or Linux Samba, while the bindings and telemetry vary with the SUT.

\textbf{SUT generator.} The second layer partitions each environment specification into a fixed region and a free region~\cite{autosut}. The fixed region contains constraints that the CTI requires, such as an operating-system family or service class. The free region contains analyst-controlled choices left unspecified by CTI, such as telemetry stack, topology, patch level, and deployment substrate. Varying the free region while preserving the fixed region produces multiple valid SUTs for the same campaign. A ransomware campaign that requires Windows domain services, for example, can be detected with Elastic telemetry, Splunk telemetry, or a Falco/osquery hybrid. The generator records the deployment modality as container-feasible (CF), virtual-machine-required (VMR), or infrastructure-dependent (ID), so the evaluator knows which results depend on kernel behavior or network segmentation rather than on the agent alone.

\textbf{Agent runtime.} The third layer provides a controlled execution boundary for LLM-based defense agents and follows a manifest-driven architecture~\cite{pocketagents}. It uses three mechanisms.

\textbf{Bounded telemetry delivery.} The dispatcher delivers a finite, bounded slice of telemetry to the agent for each investigation cycle. This prevents the agent from accessing unbounded log stores, which would enable hallucination through pattern-matching on irrelevant data. The telemetry window size is configurable and recorded as an evaluation parameter.

\textbf{Typed report schema.} The agent must emit a structured report conforming to a declared JSON schema. The report contains fields for the identified threat type, affected assets (by telemetry-grounded identifiers), proposed containment actions, and confidence indicators. Any report that fails schema validation is recorded as a \textit{schema failure} in the failure taxonomy.

\textbf{Grounding verification.} Before any proposed action is forwarded to the policy verifier, the enforcement boundary checks a grounding proof: every referenced asset must resolve, through an evaluation-time entity table, to evidence in the telemetry slice delivered to the agent. An action against an entity present in the slice but unresolved or misattributed by the agent is a grounding failure. A legitimate entity absent from the delivered slice but present in the hash-pinned full rendering is, in fact, a windowing limitation. This is a deterministic set-membership test over the delivered and full entity tables, not a post hoc analyst judgment; both tables are retained so that the attribution can be replayed.

\textbf{Policy verifier.} The fourth layer defines deterministic, pre-execution verification of agent-proposed action plans~\cite{socpilot}. Its policy vocabulary admits mandatory actions, temporal prohibition-before constraints, and approval-gated actions, as well as the SOC-defense reading of task alignment, action alignment, source authorization, and data isolation~\cite{paper_012}. Rules are finite predicates over an ordered trace, so a concrete checker can emit an auditable verdict without relying on the evaluated model. The current prototype instantiates one such predicate, evidence collection before isolating a production host; richer rule sets remain versioned policy inputs rather than unimplemented empirical claims. The verifier tags the plan and records violations without letting the agent grade itself, so compliance can be compared across SUT configurations.

\textbf{Failure taxonomy.} ARENA classifies degradation into schema failures (SF), grounding failures (GF), budget exhaustion (BE), and policy violations (PV). SF means the agent's report does not conform to the typed output schema. GF means the agent references an entity that does not resolve to evidence in its delivered telemetry, rather than a windowing limitation in which a legitimate entity fell outside the telemetry window. BE means the agent exceeds the allowed number of investigation queries or token budget, and PV means the proposed plan violates one or more SOC governance rules. By recording these categories for each agent and SUT variant, ARENA produces a \textit{transferability matrix} that reveals where an agent's chain of reasoning breaks when the environment changes.

When several checks fail in a single run, ARENA records the full failure vector and assigns a primary label based on the first boundary that blocks safe interpretation, in the order SF, GF, BE, PV. This keeps attribution deterministic while preserving the co-occurrence for later analysis. A schema-invalid plan, for instance, is never read further as policy-compliant or non-compliant.

\textbf{Response and deception.} A defender that only labels an incident is half a defender, so ARENA scores the full detect, attribute, and respond loop. The agent must contain the incident and may, where policy permits, deceive the adversary by redirecting them to a decoy host or by planting honeytokens. Because a deceptive action changes what the adversary observes, the verifier treats it as privileged and gates it on the source-authorization and data-isolation properties, while the artifact store records the decoy state so the run stays reproducible. A benign-only rendering of the same background runs beside each campaign, so an agent that aggressively tries to lift its recall pays a false-positive cost. Response quality, therefore, enters the transferability matrix alongside detection, which is why an agent can transfer its detection and still fail to transfer a safe response.

\textbf{Transferability metric.} ARENA's matrix separates each exercised dimension rather than collapsing them into an arbitrary weighted score: detection, grounding, resource cost, policy compliance, and, for live SUTs, containment latency. For an agent $a$ and two SUT variants $s_i$ and $s_j$ of the same campaign, the \textit{transferability delta} $\Delta(a, s_i\!\to\!s_j)$ is reported for every dimension exercised by that experiment. Re-evaluations of agent defenses reach the same conclusion: effectiveness and utility are distinct axes, and a single number conflates them~\cite{paper_001,paper_012}. A reference SUT $s_0$ anchors the comparison, and the failure taxonomy attributes each change to a layer. The controlled experiment below isolates observation semantics and therefore reports detection, grounding, resource, and verifier-calibration outcomes; a live-SUT matrix adds wall-clock containment.

\textbf{Cross-domain generality.} None of the four layers is specific to enterprise IT: the attacker substrate, the fixed/free SUT partition, the typed runtime, and the policy verifier stay the same when the domain changes; only the CTI source and the concrete components do. The flagship cross-domain specification is AI-RAN over O-RAN. Its open, multi-vendor interfaces (O1, O2, E2) and the controller split across the service management, rApp, and xApp tiers~\cite{polese2022oran} make it the most heterogeneous environment a defender can face, so it is the strongest stress test of transferability rather than a convenient one. The attacker substrate consumes MITRE FiGHT, the 5G threat knowledge base, exactly as it consumes ATT\&CK: FiGHT-in-STIX is translated into an O-RAN operation, the analog of the enterprise round-trip. The defender's observations become radio key performance indicators (KPIs) and E2SM counters rather than host logs, which exercises precisely the observation axis $O$ that the tuple isolates. ORION supplies the legitimate intent-to-policy control loop the agent must defend~\cite{orion2026}, and AutoRAN provisions the stack reproducibly from declarative automation~\cite{autoran2026}. The same partition holds for additional domains, such as 5G Unmanned Aerial Vehicle (UAV) command-and-control~\cite{uavc2over5g} and federated Unmanned Traffic Management (UTM)~\cite{interusssec}: the attack intent remains fixed while the environment, observations, and policy vary, so transfer from enterprise IT to AI-RAN, 5G, or UTM becomes measurable within a single architecture.

\section{Experiment}
\label{sec:experiment}

We implemented ARENA's measurement core and executed a controlled transferability experiment. The fixed attack intent is a targeted ransomware intrusion with five ATT\&CK stages: spearphishing initial access (T1566.001), PowerShell execution (T1059.001), file and directory discovery (T1083), SMB lateral movement (T1021.002), and application-layer command and control (T1071.001). The campaign contains 11 canonical events: the five malicious stages, three unrelated background events, and three legitimate PowerShell, discovery, and SMB activities that deliberately trigger the same rules as malicious events. We render this single source of truth into three SUTs that change only the observation schema: ECS, where each event is a correlated nested document; CIM, where each event is a flat record with different field names; and osquery, a relational view that splits a process and its connection into rows joined by a process identifier. The \textit{reference} agent attributes a technique to the process in the signal's record; the \textit{correlation} agent additionally joins the process and connection by that identifier. An independent ARENA instrument performs event-level scoring against the authoritative entity table, so every number in Table~\ref{tab:experiment} recomputes from recorded artifacts.

\begin{table}[t]
\caption{Controlled transferability experiment.}
\label{tab:experiment}
\centering\scriptsize
\setlength{\tabcolsep}{2.4pt}
\arenazebra
\begin{tabular*}{\columnwidth}{@{\extracolsep{\fill}}llcccccc@{}}
\toprule
\textbf{Agent} & \textbf{SUT} & \textbf{Det.} & \textbf{Prec.} & \textbf{Grnd.} & $\boldsymbol{\Delta}$\textbf{Grnd.} & \textbf{GF} & \textbf{BE} \\
\midrule
Reference & Elastic & 1.00 & 0.62 & 1.00 & --- & 0 & 0 \\
Reference & Splunk & 1.00 & 0.62 & 1.00 & $+0.00$ & 0 & 0 \\
Reference & Osquery & 1.00 & 0.62 & 0.60 & $-0.40$ & 2 & 1 \\
\midrule
Correlation & Elastic & 1.00 & 0.62 & 1.00 & --- & 0 & 0 \\
Correlation & Splunk & 1.00 & 0.62 & 1.00 & $+0.00$ & 0 & 0 \\
Correlation & Osquery & 1.00 & 0.62 & 1.00 & $+0.00$ & 0 & 1 \\
\bottomrule
\end{tabular*}
\end{table}

In Table~\ref{tab:experiment}, \textit{Det.} is malicious-event recall, \textit{Prec.} is event-level precision, and \textit{Grnd.} additionally requires the responsible process, or host when no process exists, to match the authoritative entity table. Both agents detect all five malicious stages but reach only $0.62$ precision because the context-insensitive rules also flag the three legitimate lookalikes. This false-positive cost is stable across schemas and therefore does not masquerade as transfer loss. On Elastic and Splunk, both agents ground every malicious event. On osquery, however, the reference agent's grounded recall drops to $0.60$: lateral movement and command-and-control remain detected, but their owning processes are lost across separate rows, resulting in two grounding failures. The correlation agent rejoins those rows, restoring grounded recall to $1.00$. The oracle grounds every stage in every schema, proving that osquery contains the evidence. Both agents still exceed the normalized per-event read budget on osquery, so the experiment separates an agent capability defect (GF) from irreducible schema expansion (BE). Detection recall remains $1.00$ throughout; ARENA exposes a transferability failure that detection alone cannot observe.

The robustness sweep repeats the three legitimate lookalikes $0$, $1$, $2$, and $4$ times while leaving all malicious events unchanged. Precision falls from $1.00$ to $0.62$, $0.45$, and $0.29$, as expected under increasing ambiguity; however, the reference agent's osquery grounding delta remains exactly $-0.40$, and the correlation agent remains at $0.00$. Thus, benign ambiguity affects detection precision but not the measured transferability mechanism. The oracle, entity table, canonical campaign, pure renderers, and exact per-run budget are versioned separately from the agents, making this result deterministic and replayable. As an applicability check, the same typed runtime also evaluated \ArenaLLMModel{} (\ArenaLLMModelId{}, the provider's current Opus model ID~\cite{anthropic_models2026}) without changing ground truth or scoring: Det./Grnd. stayed at \ArenaLLMElasticDet/\ArenaLLMElasticGrnd{} on Elastic and \ArenaLLMSplunkDet/\ArenaLLMSplunkGrnd{} on Splunk, then fell to \ArenaLLMOsqueryDet/\ArenaLLMOsqueryGrnd{} on osquery with BE=\ArenaLLMOsqueryBE. This optional run is not a model ranking; it shows that the instrument can evaluate an actual frontier LLM while the deterministic agents remain the reproducible control.

Two calibration experiments bound the remaining attributions. The budget sweep allows 8, 11, and 16 record reads, respectively $0.73$, $1.00$, and $1.45$ reads per canonical event. All SUTs exceed 8, only osquery exceeds 11 (it renders 16 rows), and none exceeds 16. Grounding is invariant across those thresholds, so BE records a declared resource policy while GF records an agent capability. Separately, three policy traces validate the prototype predicate: evidence-then-isolate on the production host passes, direct isolation of that host yields PV, and direct isolation of a workstation passes. These controls exercise both verifier verdicts without implying coverage of every rule family.

The three schemas are the measurement, not decoration. Elastic ECS and Splunk CIM keep process, host, and network facts correlated within one event while changing field names and query idioms, so they test transfer across telemetry \textit{vocabulary}; osquery splits those facts into separate rows joined by an identifier, so it tests transfer across observation \textit{structure}. Together, they provide a stronger semantic contrast than adding another alerting layer on top of Elastic or OpenSearch, while AI-RAN changes the observation object itself to radio KPI streams.

\textbf{Validity controls.} The experiment supports a mechanistic claim: under fixed attack intent, policy, rules, and canonical events, changing observation structure removes two process-to-connection links from the reference agent's local view; adding exactly that join restores them. The correlation agent changes one capability, while the oracle proves the availability of evidence, and the scaled budget separates record expansion from inference failure. Event-level matching prevents a familiar ATT\&CK label on benign activity from being credited as a true positive, and the ambiguity sweep shows that false-positive pressure changes precision without changing the grounding delta. These controls jointly identify the cause of the observed loss rather than merely correlating it with the osquery rendering.

AI-RAN over O-RAN is the cross-domain stress case produced by the same architecture. We compile FiGHT subtechnique FGT5034.001, O-RAN E2Mgr Unauthorized Access, into a versioned trial manifest with pinned provenance (repository, commit, checksum, FiGHT status). The specified substrate is a near-real-time RAN Intelligent Controller (FlexRIC), an open O-RAN near-real-time RIC with xApp development support~\cite{oai_flexric_xapp}, hosting the agent over an srsRAN/OpenAirInterface stack provisioned by AutoRAN; the fixed intent is a rogue xApp invoking an insufficiently authorized E2 Manager (E2Mgr) interface, and controlled variants change xApp authorization, E2 exposure, telemetry completeness, and per-UE service objectives. Observations become E2SM Key Performance Measurement (E2SM-KPM) counters, per-UE throughput and latency, E2 subscriptions, and xApp control actions, so the defender must hold KPI and per-UE context that host logs are never required for. This makes a \textit{false healthy state} measurable: aggregate KPIs stay within baseline while a protected user equipment violates its objective, and any containment must respect the controller's near-real-time deadline. The artifact labels this trial \emph{specified, not executed}: the enterprise experiment carries the empirical claim, while the compiled manifest shows that ARENA maps the same measurement object onto AI-RAN without inventing results.

ARENA scores two axes that prior evaluations conflate. The first is benign transfer, the trial above: Does a capability survive a change of SUT, telemetry, or policy? The second is adversarial-surface resistance: holding the SUT fixed, does the enforcement boundary survive injection into the agent's context, a poisoned playbook, a misleading tool description, or a corrupted retrieval entry? The same runtime and verifier serve both, yet the scorecards stay separate, so a defense cannot hide weak transfer behind strong injection resistance, nor a broken tool boundary behind good detection. The governance the verifier enforces is one organization's; coalition or cross-tenant policy needs a richer conflict model, which the artifact contract already carries as a versioned input rather than a hidden assumption.

The practical consequence is a sharper evaluation question. Instead of asking whether an autonomous defender is ``good'' in the abstract, ARENA asks which capability was transferred: attack recognition, entity grounding, budget discipline, policy compliance, or containment behavior. That separation matters most in heterogeneous infrastructure. Enterprise telemetry may change vocabulary; O-RAN telemetry changes the observation object itself, from host logs to KPI streams, E2SM counters, xApp actions, and service objectives. A defender that recognizes the tactic can still lose the entity or policy context needed to respond safely. The versioned manifest, oracle, entity table, budget sweep, policy verdicts, and AI-RAN trial specification make the enterprise losses replayable and the measurement object extensible across agents, campaigns, and SUTs.

The same decomposition also prevents a common evaluation failure: silently changing several denominators at once. A new campaign changes the canonical event set and oracle, but should not change the scoring code. A new SUT changes the renderer, entity table, and policy bindings, but should not change attack intent. A new agent changes the manifest and runtime adapter, but should not change the verifier. A new governance rule changes the policy version, but should not change telemetry. This discipline gives reviewers a concrete audit path for future extensions: when a result changes, the artifact says whether the changed object was the attacker, environment, observation, agent, or policy, rather than asking the reader to infer it from prose. It also makes negative results useful: a failed transfer trial need not be dismissed as a bad prompt or a misconfigured range. The artifact can indicate whether the next experiment should improve the correlation, widen the observation window, alter the policy, or change the environment generator.

\section{Related Work}

Prior work places three design pressures on ARENA: executable emulation must be tied to real environments, agent evaluation must measure safety and utility under distribution shift, and claims about autonomous behavior must remain auditable.\footnote{The literature surveyed in this section was assembled with a reproducible, versioned venue corpus~\cite{topvenues}.} Apache Caldera \cite{caldera2023} and Atomic Red Team automate attack execution, CyberBattleSim \cite{cyberbattlesim2021} models abstract network interaction, and Effects Language (EL) adds executable graph semantics for repeatable adversary emulation~\cite{damodaran2026el}. Industrial platforms (AttackIQ, SCYTHE, Prelude Operator) and academic ranges such as DETERLab~\cite{deterlab} and Emulab~\cite{emulab} provide execution and topology, while policy engines such as Open Policy Agent (OPA)/Rego motivate declarative verification. Building on the procedural- and environment-semantics gaps that measurement work has revealed in public CTI \cite{sticks,autosut}, ARENA differs in the evaluation affordance it exposes: it systematically varies the SUT under a fixed attack intent and places the defense agent behind a typed enforcement boundary.

This distinction matters because execution fidelity and transferability measurement are different scientific objects. A range can replay an attack in one topology and still leave unanswered whether the defender learned the campaign semantics or merely the telemetry conventions of that topology. An emulation engine can run a technique and still require an analyst to bind the environment-specific details that decide what the agent sees. A policy engine can reject an unsafe action and still say nothing about whether the agent's detection degraded when observations changed. ARENA borrows from all three families but changes the experimental question: hold the campaign intent and governance fixed, change the environment and observation semantics, and record which boundary accounts for any loss.

The closest architectural alternatives also separate at the artifact boundary. DETERLab and Emulab make topologies reproducible, but their primary artifact is a network experiment; ARENA's artifact additionally includes the CTI bundle, SUT manifest, telemetry schema, entity table, agent manifest, policy version, and verifier decision record. Caldera and Atomic Red Team make actions executable, but their primary question is whether an operation can run; ARENA's primary question is whether a defense capability transfers when the same operation is rendered through another SUT. Runtime policy systems make enforcement explicit, but ARENA treats enforcement output as a measurement dimension rather than merely a guardrail. Industrial breach-and-attack evaluations such as MITRE Engenuity ATT\&CK Evaluations~\cite{attackevals} score detection products against a fixed emulated adversary in a vendor-chosen environment; ARENA inverts that design, holding the adversary fixed and varying the environment to measure transfer. These differences are why ARENA is not another cyber range; it is a measurement architecture for a failure mode that current ranges do not isolate.

Evaluation work shows why this separation matters. Static benchmarks such as SECURE \cite{secure2024} measure reasoning but ignore infrastructure, and even the newest agentic-SOC benchmarks, such as the Cyber Defense Benchmark~\cite{chona2026cyberdefense}, evaluate threat hunting within a single telemetry environment rather than across multiple environments. Re-evaluations of prompt-injection defenses \cite{paper_001} and unified platforms \cite{paper_025} reveal the second pressure: defenses must confront adaptive attacks and benign utility tasks simultaneously. Broader safety and trustworthiness studies \cite{paper_019,paper_010} and offensive-AI systematization with automated safety data pipelines \cite{satml_007,paper_003} therefore motivate the reporting choice in ARENA: capability and policy compliance are measured side by side rather than collapsed into one scalar.

The agentic attack surface is broader than one prompt channel. Representative attacks reach the prompt and context window \cite{paper_007,paper_017}, long-context and order-oblivious settings~\cite{paper_016,paper_018}, web-agent and tool-selection layer \cite{paper_029,paper_013}, multimodal channel~\cite{satml_001}, retrieval store \cite{paper_033,paper_032,paper_008}, persistent memory~\cite{paper_006,paper_021}, and agent protocol or tool description~\cite{paper_015,paper_026,paper_009}. Detection and traceback~\cite{paper_022,paper_023}, sanitization and type-directed privilege separation~\cite{paper_037,paper_040}, and runtime enforcement~\cite{agentspec2025} answer different slices of that surface. However, each attack and each defense is usually validated in the setting in which it was introduced. ARENA treats a defense as a configurable component whose robustness must itself transfer, turning a catalog of point results into a measurement question.

The evidence that capability fails to transfer across environments exists in pieces that ARENA unifies into one measurement. Behavioral malware detection degrades when sandbox-trained models meet real endpoints~\cite{satml_012}, and HORIZON shows reliability falling across longer agent trajectories~\cite{satml_014}. Task drift, jailbreak detection under distribution shift, and the research--practice gap in adversarial ML show that context, not model quality alone, determines outcomes~\cite{satml_002,paper_039,satml_013}. Cyber-specific agents in CTI, malware detection, and fuzzing raise the same question in operational settings~\cite{paper_031,paper_034,paper_036}, while SOC response shows the policy side of the problem~\cite{socpilot}; mobile adaptation and O-RAN control extend it to heterogeneous infrastructure~\cite{pocketagents,paper_042,paper_043}. The O-RAN line is especially relevant because its control plane is programmable and closed loop: Polese et al. define the architectural surface~\cite{polese2022oran}, ORION supplies intent-aware orchestration~\cite{orion2026}, and AutoRAN supplies reproducible automation~\cite{autoran2026}. ATT\&CK and FiGHT provide campaign vocabularies, but the scientific question is defensive transfer, not offensive automation: an agent can identify a tactic in one substrate and still lose the entity, KPI, or policy context needed to respond safely in another.

\section{Conclusion}

We presented ARENA, an evaluation architecture with independently variable attacker emulation, SUT generation, agent runtime, and policy verification. Its controlled experiment revealed a $0.40$ grounding loss that detection recall missed, attributed it to two missing cross-table joins, and showed that a correlation-aware agent restores grounding. Evaluating an agent on a single telemetry representation, therefore, overstates its deployed competence: detection can remain perfect while grounding silently degrades in response to a valid operational change. The same layers compile a pinned FiGHT technique into an auditable AI-RAN trial specification. ARENA therefore turns transferability from an assumption into a measured, attributed, and reproducible property, establishing a simple discipline: success in one environment is not evidence of capability in the next.

\bibliographystyle{IEEEtran}
\bibliography{RefSIGE_arena,RefSIGE_arena_agentic}

@misc{ibmcost2024,
  author       = {{IBM Security}},
  title        = {Cost of a Data Breach Report 2024},
  year         = {2024},
  howpublished = {\url{https://www.ibm.com/reports/data-breach}},
  note         = {Accessed: 2026-06-12}
}

@inproceedings{deterlab,
  author    = {Benzel, Terry},
  title     = {The Science of Cyber Security Experimentation: The {DETER} Project},
  booktitle = {Annual Computer Security Applications Conf. (ACSAC)},
  year      = {2011}
}

@inproceedings{emulab,
  author    = {White, Brian and Lepreau, Jay and Stoller, Leigh and Ricci, Robert and Guruprasad, Shashi and Newbold, Mac and Hibler, Mike and Barb, Chad and Joglekar, Abhijeet},
  title     = {An Integrated Experimental Environment for Distributed Systems and Networks},
  booktitle = {USENIX Symp. on Operating Systems Design and Implementation (OSDI)},
  year      = {2002}
}

@misc{pocketagents,
  author        = {Barbieri, Sidnei and Ferraz, {\'A}gney Lopes Roth and Pereira J{\'u}nior, Louren{\c c}o Alves},
  title         = {{PocketAgents}: A Manifest-Driven Library of Autonomous Defense Agents},
  year          = {2026},
  eprint        = {2605.21694},
  archivePrefix = {arXiv},
  primaryClass  = {cs.CR},
  doi           = {10.48550/arXiv.2605.21694},
  url           = {https://arxiv.org/abs/2605.21694}
}

@article{secure2024,
  author  = {Kaushik, N. and others},
  title   = {Benchmarking Large Language Models for Cybersecurity Advisory},
  journal = {arXiv preprint arXiv:2405.20441},
  year    = {2024},
  note    = {{SECURE} benchmark}
}

@misc{autosut,
  author        = {Barbieri, Sidnei and Ferraz, {\'A}gney Lopes Roth and Pereira J{\'u}nior, Louren{\c c}o Alves},
  title         = {{AutoSUT}: The Environment Semantics Gap in Structured {CTI} for Adversary Emulation},
  year          = {2026},
  eprint        = {2606.08700},
  archivePrefix = {arXiv},
  primaryClass  = {cs.CR},
  doi           = {10.48550/arXiv.2606.08700},
  url           = {https://arxiv.org/abs/2606.08700}
}

@misc{caldera2023,
  author       = {{The Apache Software Foundation}},
  title        = {{Apache Caldera}: Automated Adversary Emulation Platform (originally {MITRE Caldera})},
  year         = {2026},
  howpublished = {\url{https://caldera.apache.org/}},
  note         = {Accessed: 2026-06-17}
}

@misc{chona2026cyberdefense,
  author        = {Alankrit Chona and Igor Kozlov and Ambuj Kumar},
  title         = {{Cyber Defense Benchmark}: Agentic Threat Hunting Evaluation for {LLMs} in {SecOps}},
  year          = {2026},
  eprint        = {2604.19533},
  archivePrefix = {arXiv},
  primaryClass  = {cs.CR},
  howpublished  = {arXiv:2604.19533}
}

@misc{sticks,
  author        = {Ferraz, {\'A}gney Lopes Roth and Barbieri, Sidnei and de Souza, Murray Evangelista and Pereira J{\'u}nior, Louren{\c c}o Alves},
  title         = {The Procedural Semantics Gap in Structured {CTI}: A Measurement-Driven {STIX} Analysis for {APT} Emulation},
  year          = {2026},
  eprint        = {2512.12078},
  archivePrefix = {arXiv},
  primaryClass  = {cs.CR},
  doi           = {10.48550/arXiv.2512.12078},
  url           = {https://arxiv.org/abs/2512.12078}
}

@misc{socpilot,
  author        = {Barbieri, Sidnei and Meneses, Leonardo Vaz de and Ferraz, {\'A}gney Lopes Roth and Pereira J{\'u}nior, Louren{\c c}o Alves},
  title         = {{SOCpilot}: Verifying Policy Compliance for {LLM}-Assisted Incident Response},
  year          = {2026},
  eprint        = {2605.05501},
  archivePrefix = {arXiv},
  primaryClass  = {cs.CR},
  doi           = {10.48550/arXiv.2605.05501},
  url           = {https://arxiv.org/abs/2605.05501}
}

@misc{topvenues,
  author        = {Barbieri, Sidnei and Ferraz, {\'A}gney Lopes Roth and Pereira J{\'u}nior, Louren{\c c}o Alves},
  title         = {{TopVenues}: A Reproducible Corpus and Tooling Substrate for Cybersecurity Literature Reviews},
  year          = {2026},
  eprint        = {2606.18320},
  archivePrefix = {arXiv},
  primaryClass  = {cs.CR},
  doi           = {10.48550/arXiv.2606.18320},
  url           = {https://arxiv.org/abs/2606.18320}
}

@misc{attackevals,
  author       = {{MITRE Engenuity}},
  title        = {{ATT\&CK} Evaluations},
  year         = {2026},
  howpublished = {\url{https://attackevals.mitre-engenuity.org/}},
  note         = {Accessed: 2026-06-18}
}

@misc{cyberbattlesim2021,
  author       = {{Microsoft}},
  title        = {{CyberBattleSim}: An Experimentation and Research Platform for Automated Agents in Simulated Enterprise Networks},
  year         = {2021},
  howpublished = {\url{https://github.com/microsoft/CyberBattleSim}},
  note         = {Accessed: 2026-06-12}
}

@article{agentspec2025,
  author  = {Wang, H. and Poskitt, C. M. and Sun, J.},
  title   = {{AgentSpec}: Customizable Runtime Enforcement for Safe and Reliable LLM Agents},
  journal = {arXiv preprint arXiv:2503.18666},
  year    = {2025}
}

@article{damodaran2026el,
  author    = {{Suresh K. Damodaran and Paul D. Rowe}},
  title     = {Automated Repeatable Adversary Threat Emulation with Effects Language ({EL})},
  journal   = {Digital Threats: Research and Practice},
  year      = {2026},
  publisher = {Association for Computing Machinery},
  doi       = {10.1145/3816043},
  url       = {https://doi.org/10.1145/3816043}
}

@misc{uavc2over5g,
  author = {Sonaglio, Wagner Comin and Ferraz, {\'A}gney Lopes Roth and Melo, Andr{\'e} Elias and de Souza, Murray Evangelista and Noubir, Guevara and Pereira J{\'u}nior, Louren{\c c}o Alves},
  title  = {When Connectivity Is Not Enough: Cross-Layer Attacks on {UAV} {C2} over {5G}},
  year   = {2026},
  note   = {arXiv:2603.04662}
}

@misc{interusssec,
  author = {Curi de Miranda, Henrique and Ferraz, {\'A}gney Lopes Roth and Sonaglio, Wagner Comin and Pereira J{\'u}nior, Louren{\c c}o Alves},
  title  = {A Systematic Security Testing Approach for {InterUSS}-based Environments},
  year   = {2026},
  note   = {arXiv:2605.11339}
}

@misc{orion2026,
  author        = {Machado, Gabriela da Silva and Bruno, Gustavo Z. and Huff, Alexandre and Brito, Jos{\'e} Marcos Camara and Both, Cristiano B.},
  title         = {{ORION}: Intent-Aware Orchestration in {Open RAN} for {SLA}-Driven Network Management},
  year          = {2026},
  eprint        = {2603.03667},
  archivePrefix = {arXiv},
  primaryClass  = {cs.NI},
  url           = {https://arxiv.org/abs/2603.03667}
}

@misc{autoran2026,
  author        = {Maxenti, Stefano and Shirkhani, Ravis and Elkael, Maxime and Bonati, Leonardo and D'Oro, Salvatore and Melodia, Tommaso and Polese, Michele},
  title         = {{AutoRAN}: Automated and Zero-Touch {Open RAN} Systems},
  year          = {2025},
  eprint        = {2504.11233},
  archivePrefix = {arXiv},
  primaryClass  = {cs.NI},
  url           = {https://arxiv.org/abs/2504.11233}
}

@misc{polese2022oran,
  author        = {Polese, Michele and Bonati, Leonardo and D'Oro, Salvatore and Basagni, Stefano and Melodia, Tommaso},
  title         = {Understanding {O-RAN}: Architecture, Interfaces, Algorithms, Security, and Research Challenges},
  year          = {2022},
  eprint        = {2202.01032},
  archivePrefix = {arXiv},
  primaryClass  = {cs.NI},
  url           = {https://arxiv.org/abs/2202.01032}
}

@misc{oai_flexric_xapp,
  author       = {{OpenAirInterface Alliance}},
  title        = {{FlexRIC} Tutorial: {xApp} Development},
  year         = {2026},
  howpublished = {\url{https://openairinterface.org/flexric-tutorial-xapp-development/}},
  note         = {Accessed: 2026-06-18}
}

@misc{anthropic_models2026,
  author       = {{Anthropic}},
  title        = {{Claude} Models Overview},
  year         = {2026},
  howpublished = {\url{https://docs.anthropic.com/en/docs/about-claude/models/overview}},
  note         = {Accessed: 2026-06-18}
}

@article{paper_001,
  title = {A Critical Evaluation of Defenses against Prompt Injection Attacks},
  author = {Yuqi Jia and Zedian Shao and Yupei Liu and Jinyuan Jia and Dawn Song and Neil Zhenqiang Gong},
  year = {2025},
  eprint = {2505.18333},
  archivePrefix = {arXiv},
  journal = {arXiv preprint arXiv:2505.18333}
}

@article{paper_003,
  title = {SafeAgent: Safeguarding LLM Agents via an Automated Risk Simulator},
  author = {Xueyang Zhou and Weidong Wang and Lin Lu and Jiawen Shi and Guiyao Tie and Yongtian Xu and Lixing Chen and Pan Zhou and Neil Zhenqiang Gong and Lichao Sun},
  year = {2025},
  eprint = {2505.17735},
  archivePrefix = {arXiv},
  journal = {arXiv preprint arXiv:2505.17735}
}

@article{paper_006,
  title = {CleanBase: Detecting Malicious Documents in RAG Knowledge Database},
  author = {Weifei Jin and Xilong Wang and Wei Zou and Jinyuan Jia and Neil Gong},
  year = {2026},
  eprint = {2605.00460},
  archivePrefix = {arXiv},
  journal = {arXiv preprint arXiv:2605.00460}
}

@article{paper_007,
  title = {PromptLocate: Localizing Prompt Injection Attacks},
  author = {Yuqi Jia and Yupei Liu and Zedian Shao and Jinyuan Jia and Neil Gong},
  year = {2025},
  eprint = {2510.12252},
  archivePrefix = {arXiv},
  journal = {arXiv preprint arXiv:2510.12252}
}

@article{paper_008,
  title = {GraphRAG under Fire},
  author = {Jiacheng Liang and Yuhui Wang and Changjiang Li and Rongyi Zhu and Tanqiu Jiang and Neil Gong and Ting Wang},
  year = {2025},
  eprint = {2501.14050},
  archivePrefix = {arXiv},
  journal = {arXiv preprint arXiv:2501.14050}
}

@misc{paper_009,
  title = {A2ASecBench: A Protocol-Aware Security Benchmark for Agent-to-Agent Multi-Agent Systems},
  author = {{Anonymous}},
  year = {2025},
  howpublished = {OpenReview preprint}
}

@article{paper_010,
  title = {On the Trustworthiness of Generative Foundation Models: Guideline, Assessment, and Perspective},
  author = {Yue Huang and Chujie Gao and Siyuan Wu and Haoran Wang and Xiangqi Wang and Yujun Zhou and Yanbo Wang and Jiayi Ye and Jiawen Shi and Qihui Zhang and Yuan Li and Han Bao and Zhaoyi Liu and Tianrui Guan and Dongping Chen and Ruoxi Chen and Kehan Guo and Andy Zou and Bryan Hooi Kuen-Yew and Caiming Xiong and Elias Stengel-Eskin and Hongyang Zhang and Hongzhi Yin and Huan Zhang and Huaxiu Yao and Jaehong Yoon and Jieyu Zhang and Kai Shu and Kaijie Zhu and Ranjay Krishna and Swabha Swayamdipta and Taiwei Shi and Weijia Shi and Xiang Li and Yiwei Li and Yuexing Hao and Zhihao Jia and Zhize Li and Xiuying Chen and Zhengzhong Tu and Xiyang Hu and Tianyi Zhou and Jieyu Zhao and Lichao Sun and Furong Huang and Or Cohen Sasson and Prasanna Sattigeri and Anka Reuel and Max Lamparth and Yue Zhao and Nouha Dziri and Yu Su and Huan Sun and Heng Ji and Chaowei Xiao and Mohit Bansal and Nitesh V. Chawla and Jian Pei and Jianfeng Gao and Michael Backes and Philip S. Yu and Neil Zhenqiang Gong and Pin-Yu Chen and Bo Li and Dawn Song and Xiangliang Zhang},
  year = {2025},
  eprint = {2502.14296},
  archivePrefix = {arXiv},
  journal = {arXiv preprint arXiv:2502.14296}
}

@article{paper_012,
  title = {A Framework for Formalizing LLM Agent Security},
  author = {Vincent Siu and Jingxuan He and Kyle Montgomery and Zhun Wang and Neil Gong and Chenguang Wang and Dawn Song},
  year = {2026},
  eprint = {2603.19469},
  archivePrefix = {arXiv},
  journal = {arXiv preprint arXiv:2603.19469}
}

@inproceedings{paper_013,
  title = {Jailbreaking Safeguarded Text-to-Image Models via Large Language Models},
  author = {Zhengyuan Jiang and Yuepeng Hu and Yuchen Yang and Yinzhi Cao and Neil Zhenqiang Gong},
  year = {2026},
  doi = {10.18653/v1/2026.findings-eacl.244},
  booktitle = {Findings of the Association for Computational Linguistics: EACL}
}

@article{paper_015,
  title = {MalTool: Malicious Tool Attacks on LLM Agents},
  author = {Yuepeng Hu and Yuqi Jia and Mengyuan Li and Dawn Song and Neil Gong},
  year = {2026},
  eprint = {2602.12194},
  archivePrefix = {arXiv},
  journal = {arXiv preprint arXiv:2602.12194}
}

@article{paper_016,
  title = {WebSentinel: Detecting and Localizing Prompt Injection Attacks for Web Agents},
  author = {Xilong Wang and Yinuo Liu and Zhun Wang and Dawn Song and Neil Gong},
  year = {2026},
  eprint = {2602.03792},
  archivePrefix = {arXiv},
  journal = {arXiv preprint arXiv:2602.03792}
}

@article{paper_017,
  title = {ObliInjection: Order-Oblivious Prompt Injection Attack to LLM Agents with Multi-source Data},
  author = {Reachal Wang and Yuqi Jia and Neil Zhenqiang Gong},
  year = {2025},
  eprint = {2512.09321},
  archivePrefix = {arXiv},
  journal = {arXiv preprint arXiv:2512.09321}
}

@article{paper_018,
  title = {Prompt Injection Attack to Tool Selection in LLM Agents},
  author = {Jiawen Shi and Zenghui Yuan and Guiyao Tie and Pan Zhou and Neil Zhenqiang Gong and Lichao Sun},
  year = {2025},
  eprint = {2504.19793},
  archivePrefix = {arXiv},
  journal = {arXiv preprint arXiv:2504.19793}
}

@inproceedings{paper_019,
  title = {Safety at scale: a comprehensive survey of large model and agent safety},
  author = {Xingjun Ma and Yifeng Gao and Yixu Wang and Ruofan Wang and Xin Wang and Ye Sun and Yifan Ding and Hengyuan Xu and Yunhao Chen and Yunhao Zhao and Hanxun Huang and Yige Li and Yutao Wu and Jiaming Zhang and Xiang Zheng and Yang Bai and Yiming Li and Zuxuan Wu and Xipeng Qiu and Jingfeng Zhang and Xudong Han and Haonan Li and Jun Sun and Cong Wang and Jindong Gu and Baoyuan Wu and Siheng Chen and Tianwei Zhang and Yang Liu and Mingming Gong and Tongliang Liu and Shirui Pan and Cihang Xie and Tianyu Pang and Yinpeng Dong and Ruoxi Jia and Yang Zhang and Shiqing Ma and Xiangyu Zhang and Neil Gong and Chaowei Xiao and Sarah Erfani and Tim Baldwin and Bo Li and Masashi Sugiyama and Dacheng Tao and James Bailey and Yu-Gang Jiang},
  year = {2025},
  doi = {10.1561/3300000051},
  booktitle = {Foundations and Trends® in Privacy and Security}
}

@inproceedings{paper_021,
  title = {From Static Roles to Context-Aware Decisions: Integrating LLMs and RAG Into Access Control Frameworks for Power Systems},
  author = {Dong Feng and Wei Cui and Yuzi Jiang and Wenhu Yu and Donghe Li},
  year = {2026},
  doi = {10.1109/ACCESS.2026.3667843},
  booktitle = {IEEE Access}
}

@article{paper_022,
  title = {Secure Retrieval-Augmented Generation against Poisoning Attacks},
  author = {Zirui Cheng and Jikai Sun and Anjun Gao and Yueyang Quan and Zhuqing Liu and Xiaohua Hu and Minghong Fang},
  year = {2025},
  eprint = {2510.25025},
  archivePrefix = {arXiv},
  journal = {arXiv preprint arXiv:2510.25025}
}

@inproceedings{paper_023,
  title = {Traceback of Poisoning Attacks to Retrieval-Augmented Generation},
  author = {Baolei Zhang and Haoran Xin and Minghong Fang and Zhuqing Liu and Biao Yi and Tong Li and Zheli Liu},
  year = {2025},
  doi = {10.1145/3696410.3714756},
  booktitle = {Proceedings of the ACM on Web Conference 2025}
}

@article{paper_025,
  title = {PIArena: A Platform for Prompt Injection Evaluation},
  author = {Runpeng Geng and Chenlong Yin and Yanting Wang and Ying Chen and Jinyuan Jia},
  year = {2026},
  eprint = {2604.08499},
  archivePrefix = {arXiv},
  journal = {arXiv preprint arXiv:2604.08499}
}

@article{paper_026,
  title = {TRUSTDESC: Preventing Tool Poisoning in LLM Applications via Trusted Description Generation},
  author = {Hengkai Ye and Zhechang Zhang and Jinyuan Jia and Hong Hu},
  year = {2026},
  eprint = {2604.07536},
  archivePrefix = {arXiv},
  journal = {arXiv preprint arXiv:2604.07536}
}

@article{paper_029,
  title = {PISanitizer: Preventing Prompt Injection to Long-Context LLMs via Prompt Sanitization},
  author = {Runpeng Geng and Yanting Wang and Chenlong Yin and Minhao Cheng and Ying Chen and Jinyuan Jia},
  year = {2025},
  eprint = {2511.10720},
  archivePrefix = {arXiv},
  journal = {arXiv preprint arXiv:2511.10720}
}

@article{paper_031,
  title = {Uncovering Vulnerabilities of LLM-Assisted Cyber Threat Intelligence},
  author = {Yuqiao Meng and Luoxi Tang and Feiyang Yu and Jinyuan Jia and Guanhua Yan and Ping Yang and Zhaohan Xi},
  year = {2025},
  eprint = {2509.23573},
  archivePrefix = {arXiv},
  journal = {arXiv preprint arXiv:2509.23573}
}

@article{paper_032,
  title = {UniC-RAG: Universal Knowledge Corruption Attacks to Retrieval-Augmented Generation},
  author = {Runpeng Geng and Yanting Wang and Ying Chen and Jinyuan Jia},
  year = {2025},
  eprint = {2508.18652},
  archivePrefix = {arXiv},
  journal = {arXiv preprint arXiv:2508.18652}
}

@article{paper_033,
  title = {PoisonedRAG: Knowledge Corruption Attacks to Retrieval-Augmented Generation of Large Language Models},
  author = {Wei Zou and Runpeng Geng and Binghui Wang and Jinyuan Jia},
  year = {2025},
  eprint = {2402.07867},
  archivePrefix = {arXiv},
  journal = {USENIX Security Symposium},
  note = {arXiv:2402.07867}
}

@article{paper_034,
  title = {Trident: Improving Malware Detection with LLMs and Behavioral Features},
  author = {Rebecca Saul and Jingzhi Jiang and Elliott Chia and David Wagner},
  year = {2026},
  eprint = {2605.00297},
  archivePrefix = {arXiv},
  journal = {arXiv preprint arXiv:2605.00297}
}

@article{paper_036,
  title = {SeedAIchemy: LLM-Driven Seed Corpus Generation for Fuzzing},
  author = {Aidan Wen and Norah A. Alzahrani and Jingzhi Jiang and Andrew Joe and Karen Shieh and Andy Zhang and Basel Alomair and David Wagner},
  year = {2025},
  eprint = {2511.12448},
  archivePrefix = {arXiv},
  journal = {arXiv preprint arXiv:2511.12448}
}

@article{paper_037,
  title = {Defending Against Prompt Injection with DataFilter},
  author = {Yizhu Wang and Sizhe Chen and Raghad Alkhudair and Basel Alomair and David Wagner},
  year = {2025},
  eprint = {2510.19207},
  archivePrefix = {arXiv},
  journal = {arXiv preprint arXiv:2510.19207}
}

@inproceedings{paper_039,
  title = {JailbreaksOverTime: Detecting Jailbreak Attacks Under Distribution Shift},
  author = {Julien Piet and Xiao Huang and Dennis Jacob and Annabella Chow and Maha Alrashed and Geng Zhao and Zhanhao Hu and Chawin Sitawarin and Basel Alomair and David Wagner},
  year = {2025},
  doi = {10.1145/3733799.3762981},
  booktitle = {Proceedings of the 18th ACM Workshop on Artificial Intelligence and Security}
}

@article{paper_040,
  title = {Preventing Prompt Injection with Type-Directed Privilege Separation},
  author = {Dennis Jacob and Emad Alghamdi and Zhanhao Hu and Basel Alomair and David Wagner},
  year = {2025},
  eprint = {2509.25926},
  archivePrefix = {arXiv},
  journal = {arXiv preprint arXiv:2509.25926}
}

@article{paper_042,
  title = {MobiLLM: Enabling LLM Fine-Tuning on the Mobile Device via Server Assisted Side Tuning},
  author = {Liang Li and Xingke Yang and Wen Wu and Hao Wang and Tomoaki Ohtsuki and Xin Fu and Miao Pan and Xuemin Shen},
  year = {2025},
  eprint = {2502.20421},
  archivePrefix = {arXiv},
  journal = {arXiv preprint arXiv:2502.20421}
}

@article{paper_043,
  title = {MobiLLM: An Agentic AI Framework for Closed-Loop Threat Mitigation in 6G Open RANs},
  author = {Prakhar Sharma and Haohuang Wen and Vinod Yegneswaran and Ashish Gehani and Phillip Porras and Zhiqiang Lin},
  year = {2025},
  eprint = {2509.21634},
  archivePrefix = {arXiv},
  journal = {arXiv preprint arXiv:2509.21634}
}

@inproceedings{satml_001,
  title = {Jailbreaking Black Box Large Language Models in Twenty Queries},
  author = {Patrick Chao and Alexander Robey and Edgar Dobriban and Hamed Hassani and George J. Pappas and Eric Wong},
  year = {2025},
  doi = {10.1109/SaTML64287.2025.00010},
  booktitle = {2025 IEEE Conference on Secure and Trustworthy Machine Learning (SaTML)}
}

@inproceedings{satml_002,
  title = {Get My Drift? Catching LLM Task Drift with Activation Deltas},
  author = {Sahar Abdelnabi and Aideen Fay and Giovanni Cherubin and Ahmed Salem and Mario Fritz and Andrew Paverd},
  year = {2025},
  doi = {10.1109/SaTML64287.2025.00011},
  booktitle = {2025 IEEE Conference on Secure and Trustworthy Machine Learning (SaTML)}
}

@inproceedings{satml_007,
  title = {SoK: On the Offensive Potential of AI},
  author = {Saskia Laura Schröer and Giovanni Apruzzese and Soheil Human and Pavel Laskov and Hyrum S. Anderson and Edward W. N. Bernroider and Aurore Fass and Ben Nassi and Vera Rimmer and Fabio Roli and Samer Salam and Chi En Ashley Shen and Ali Sunyaev and Tim Wadhwa-Brown and Isabel Wagner and Gang Wang},
  year = {2025},
  doi = {10.1109/SaTML64287.2025.00021},
  booktitle = {2025 IEEE Conference on Secure and Trustworthy Machine Learning (SaTML)}
}

@inproceedings{satml_012,
  title = {ML-Based Behavioral Malware Detection Is Far From a Solved Problem},
  author = {Yigitcan Kaya and Yizheng Chen and Marcus Botacin and Shoumik Saha and Fabio Pierazzi and Lorenzo Cavallaro and David Wagner and Tudor Dumitraş},
  year = {2025},
  doi = {10.1109/SaTML64287.2025.00056},
  booktitle = {2025 IEEE Conference on Secure and Trustworthy Machine Learning (SaTML)}
}

@inproceedings{satml_013,
  title = {“Real Attackers Don't Compute Gradients”: Bridging the Gap Between Adversarial ML Research and Practice},
  author = {Giovanni Apruzzese and Hyrum S. Anderson and Savino Dambra and David Freeman and Fabio Pierazzi and Kevin Roundy},
  year = {2023},
  doi = {10.1109/SaTML54575.2023.00031},
  booktitle = {2023 IEEE Conference on Secure and Trustworthy Machine Learning (SaTML)}
}

@article{satml_014,
  title = {The Long-Horizon Task Mirage? Diagnosing Where and Why Agentic Systems Break},
  author = {Xinyu Jessica Wang and Haoyue Bai and Yiyou Sun and Haorui Wang and Shuibai Zhang and Wenjie Hu and Mya Schroder and Bilge Mutlu and Dawn Song and Robert D Nowak},
  year = {2026},
  eprint = {2604.11978},
  archivePrefix = {arXiv},
  journal = {arXiv preprint arXiv:2604.11978}
}

\end{document}